\documentclass[a4paper,12pt]{article}
\usepackage{graphicx,rotating,colortbl,xcolor,wrapfig,hyperref,slashed,cite}
\usepackage{amsmath,bbold,mathrsfs,amssymb}

\hypersetup{colorlinks,bookmarksopen,bookmarksnumbered,
linkcolor=blus,pdfstartview=FitH,urlcolor=rossos,citecolor=verde}

\newcommand{\h}{\phi}
 \newcommand{\fig}[1]{~\ref{fig:#1}}
\newcommand{\ggr}{g_{\rm grav}}

\newcommand{\xxx}[1]{{\color{magenta}[\bf #1]}}

\def\to{\rightarrow}
\def\bea{\begin{eqnarray}}
\def\eea{\end{eqnarray}}
\def\eq#1{~(\ref{eq:#1})}

\newcommand{\sfrac}[2]{{#1/#2}}
\definecolor{Gray}{gray}{0.95}

\definecolor{rosso}{cmyk}{0,1,1,0.4}
\definecolor{rossos}{cmyk}{0,1,1,0.55}
\definecolor{rossoc}{cmyk}{0,1,1,0.2}
\definecolor{blu}{cmyk}{1,1,0,0.3}
\definecolor{blus}{cmyk}{1,1,0,0.6}
\definecolor{bluc}{cmyk}{1,1,0,0.1}
\definecolor{verde}{cmyk}{0.92,0,0.59,0.25}
\definecolor{verdec}{cmyk}{0.92,0,0.59,0.15}
\definecolor{verdes}{cmyk}{0.92,0,0.59,0.4}

\newcommand{\eV}{\,{\rm eV}}
\newcommand{\SU}{\,{\rm SU}}

 \def\be   {\begin{equation}}   \def\ee   {\end{equation}}
 \def\ba   {\begin{array}}      \def\ea   {\end{array}}

\def\Lag{\mathscr{L}}

\newcommand{\bp}{\bar M_{\rm Pl}}
 \newcommand{\med}[1]{\langle #1\rangle}
\makeatletter
\newcounter{alphaequation}[equation]
\def\thealphaequation{\theequation\hbox to
0.6em{\hfil\alph{alphaequation}\hfil}}
\def\eqnsystem#1{
\def\@eqnnum{{\rm (\thealphaequation)}}
\def\@@eqncr{\let\@tempa\relax \ifcase\@eqcnt \def\@tempa{& & &} \or
  \def\@tempa{& &}\or \def\@tempa{&}\fi\@tempa
  \if@eqnsw\@eqnnum\refstepcounter{alphaequation}\fi
\global\@eqnswtrue\global\@eqcnt=0\cr}
\refstepcounter{equation} \let\@currentlabel\theequation \def\@tempb{#1}
\ifx\@tempb\empty\else\label{#1}\fi
\refstepcounter{alphaequation}
\let\@currentlabel\thealphaequation
\global\@eqnswtrue\global\@eqcnt=0 \tabskip\@centering\let\\=\@eqncr
$$\halign to \displaywidth\bgroup \@eqnsel\hskip\@centering
$\displaystyle\tabskip\z@{##}$&\global\@eqcnt\@ne
\hskip2\arraycolsep\hfil${##}$\hfil& \global\@eqcnt\tw@\hskip2\arraycolsep
$\displaystyle\tabskip\z@{##}$\hfil
\tabskip\@centering&\llap{##}\tabskip\z@\cr}
\def\endeqnsystem{\@@eqncr\egroup$$\global\@ignoretrue} \makeatother

\newcommand{\GeV}{\,{\rm GeV}}

\newcommand{\Tr}{\,{\rm Tr}}
\newcommand{\diag}{\,{\rm diag}}

\def\circa#1{\,\raise.3ex\hbox{$#1$\kern-.75em\lower1ex\hbox{$\sim$}}\,}

\newcommand{\beq}{\begin{equation}}
\newcommand{\eeq}{\end{equation}}

\font\ital=cmu10 

\makeatletter
\def\hhref#1{\href{http://arxiv.org/abs/#1}{arXiv:#1}}
\usepackage{xstring} 
\newcommand{\hhrefq}[1]{\IfSubStr{#1}{:}{\href{http://inspirehep.net/search?ln=en&ln=en&p=#1&of=hb&action_search=Search&sf=&so=d&rm=&rg=25&sc=0}{InSpire:#1}}{\hhref{#1}}}

\def\art{\@ifnextchar[{\eart}{\oart}}
\def\eart[#1]#2#3#4#5#6{{\rm #2}, {\em #3 \bf #4} {\rm (#6) #5} ({\em #1})}
\def\article{\@ifnextchar[{\earticle}{\oarticle}}
\def\oarticle#1#2#3#4#5#6{{\rm #1}, {\ital ``#6''}, {\rm #2 #3 (#5) #4}}
\def\earticle[#1]#2#3#4#5#6#7{{\rm #2}, {\ital ``#7''}, {\rm #3 #4 (#6) #5}  [\hhrefq{#1}]}
\def\hepart[#1]#2{{\rm #2, \ital#1}}
\def\heparticle[#1]#2#3{#2, {\ital ``#3''} [\hhrefq{#1}]}
\newcommand{\doi}[1]{\href{http://dx.doi.org/#1}{[link]}}

\def\hhref#1{\href{http://arxiv.org/abs/#1}{arXiv:#1}} 

\begin{document}
\thispagestyle{empty}
\begin{center}
{\LARGE \bf \color{rossos} A landscape for the cosmological \\ constant and the Higgs mass}  \\[5mm]
{\bf Parsa Ghorbani$^{a,b}$},  {\bf Alessandro Strumia$^{a}$}, {\bf Daniele Teresi$^{a,b}$}\\[3mm]
{\it $^a$ Dipartimento di Fisica dell'Universit{\`a} di Pisa}\\[1mm]
{\it $^b$ INFN, Sezione di Pisa, Italy}\\[1mm]
\vspace{0.5cm}
{\large\bf\color{blus} Abstract}
\begin{quote}
The cosmological constant and the Higgs mass seem unnaturally small and anthropically selected. 
We show that both can be efficiently scanned in
Quantum Field Theories with a large enough number of vacua controllable thanks to
approximated $\mathbb{Z}_2$ symmetries 
(even for Coleman--Weinberg potentials).
We find that vacuum decay in a landscape implies 
weaker bounds than previously estimated.
Special vacua where one light scalar is accidentally light
avoid catastrophic vacuum decay if its self-cubic is absent.
This is what happens for the Higgs doublet, thanks to gauge invariance.
Yukawa couplings can be efficiently scanned,
as suggested by anthropic boundaries on light quark masses.
Finally, we suggest that the lack of predictivity of landscapes can be mitigated if
their probability distributions are non-Gaussian (possibly even fractal). 
\end{quote}
\end{center}

{
\tableofcontents}
\normalsize

\section{Introduction}
The measured vacuum energy $V\approx (2.3~10^{-3}\eV)^4$ is unnaturally small
in the Standard Model.
The measured weak scale $v \approx 174\GeV$ is unnaturally small
in extensions of the Standard Model with heavy new physics significantly coupled to the Higgs, 
for example in string theory.
Motivated by these naturalness issues,
theorists tried to devise natural extensions of the Standard Model such that
new physics keeps $V$ and $v$ naturally small.
But such new physics has not been observed and experimental bounds are now so strong 
that proposed solutions become tuned and/or theoretically contrived.

On the other hand, it has been noticed that the smallness of both $V$ and $v$
can be interpreted in terms of anthropic selection --- a topic
started by astrophysicists~\cite{Carter:1974zz,Carr:1979sg,Barrow:1988yia} that acquired
relevance for fundamental theory.
Values of $V$ and $v$ much smaller than the Planck scale are generically needed to have big objects within horizons.
What is striking is that the measured values of $V$ and $v$ as well as of light fermion masses
$m_f$
seem close to plausible anthropic boundaries:
\begin{itemize}
\item Galaxies can only form if the
vacuum energy  is at most two orders of magnitude larger than its observed value~\cite{Weinberg:1988cp}.  
This argument was originally used by Weinberg to expect a vacuum energy
at a level around what has been later observed.

\item
Weak interactions play a major role 
only in two cosmological circumstances:
Big Bang Nucleosynthesis (see~\cite{1409.0551}) and core-collapse supernova explosions~\cite{1906.00986}.
In both cases special physics happens for $v \sim \Lambda_{\rm QCD}^{3/4} M_{\rm Pl}^{1/4}$,
up to omitted order one factors such that the critical value of $v$ is close to the measured value.
In this special small range of $v$
neutrino-driven core-collapse supernova explosions spread
intermediate-mass nuclei presumably necessary for `life', 
such that weak interactions might be playing a role of anthropic relevance~\cite{1906.00986}.

\item
Furthermore, the values of light fermion masses $m_u$, $m_d$, $m_e$
lie not far from anthropic boundaries~\cite{hep-ph/9707380,0712.2454,0809.1647}.
For example, 
the Standard Model allows for a complex chemistry of
many nuclei because the proton is the lightest baryon,
and neutrons are stable within nuclei (while being unstable as free  particles):
these and other features arise thanks to quark masses with appropriate numerical values.
In the Standard Model $m_f = y_f v$: anthropic selection of $m_f$
would imply an extra indirect anthropic bound on the weak scale $v$ 
if Yukawa couplings $y_f$ had fixed values in the landscape
(a possibility that seems unlikely in view of the multiple boundaries on $m_u$, $m_d$, $m_e$). 

\end{itemize}
These findings indicate that anthropic selection seems playing a role.
Anthropic selection is realised, compatibly 
with our present observational and theoretical understanding,
in theories that possess a huge number $N \gg 10^{160}$
of physically inequivalent local minima. 
Then, inflationary dynamics and vacuum decay can give rise to a `multiverse'
with large super-horizon regions of space in different local minima. 

Theories where the smallness of $v,V$ is attributed to
anthropic selection received much less attention
than theories based on symmetries or other natural mechanisms.
Indeed physicists and philosophers prefer
theories that are predictive and testable.
An unique mono-vacuist  theory  able of predict everything
is physicists' dream, but Nature might have chosen otherwise: 
a poly-vacuistic nightmare
that would downgrade our present observational
understanding of physics to stamp collecting with one stamp,
as experiments in our universe might not allow to test other vacua.
On the other hand, available experimental data disfavoured so much 
falsifiable natural theories that the falsifiability criterion itself prompts us to consider
the multiverse hypothesis.

\medskip

Proceeding in this direction, basic multiverse issues have not yet been considered much.
Can a large enough and diverse enough number of vacua 
be obtained within renormalizable relativistic quantum field theory (QFT) in 3+1 dimensions
where scalar potentials are at most quartic?
Or a multiverse needs some deeper theory, such as string theory,
that reduces to QFT as its low-energy Taylor-like expansion
(or, equivalently, large-distance multipole-like expansion)?

In general, the vacuum energy $V$ (and possibly the weak scale $v$) 
receive dominant contributions from the heaviest states of the theory,
so that any QFT description of the landscape might be incomplete.
We proceed with a QFT description assuming  that heavier states
(if present) do not dominantly contribute to $V$ and $v$,
for example because of supersymmetry broken in appropriate ways.
%
%
%

A concrete problem is that anthropic selection needs
many more vacua that what can be counted in practice.
For example, string theory might give rise to a complicated landscape,
but its existence is not  established.
In order to know that a big landscape of vacua really exists, one
needs to assume landscapes with special controllable structures.
Finding vacua with small vacuum energy is another
possibly NP-hard problem~\cite{hep-th/0602072}.

A special QFT landscape for the cosmological constant was obtained using
a few hundreds of {\em independent} scalars $\h_i$
with two non-degenerate minima each, giving $2^N$ different values of the vacuum energy~\cite{NimaLand}.
However, a landscape of independent scalars does not generate a wide range of values for the
Higgs mass, unless the Higgs boson is introduced as a special field coupled to the others.
A landscape where the Higgs is one among many scalars can be problematic:
mixed scalar interactions destroy the landscape 
(drastically reducing the number of vacua, as was generically estimated
using random matrix theory~\cite{hep-th/0512050})
around when they are big enough that scalar masses are efficiently scanned,
as needed  for an  anthropic interpretation of the weak scale.

Does a landscape for the cosmological constant and for the Higgs mass exist?

In section~\ref{QFTland} we will answer positively, considering a different landscape, 
where an approximate $\mathbb{Z}_2^N$ symmetry
makes computations tractable.
We find that mixed quartics can be as large as scalar self-quartics, and that, under reasonable conditions,  the vacuum energy and the squared Higgs mass and Yukawa couplings can be efficiently scanned.
In section~\ref{vac} we reconsider vacuum stability, finding weaker bounds than in~\cite{NimaLand},
and that special vacua where one light scalar is accidentally light
avoid catastrophic vacuum decay if its self-cubic is absent.
This is what happens for the Higgs doublet, thanks to gauge invariance.
In section~\ref{CW} we show that Coleman-Weinberg potentials
are not incompatible with a landscape.
In section~\ref{nonGauss} we consider a non-Gaussian
landscape probability distribution that might result in enhanced predictivity.
Conclusions are  given in section~\ref{Conclusions}.

\section{A Quantum Field Theory landscape?}\label{QFTland}

We consider a QFT in $3+1$ space-time dimensions with $N$ real scalars $\h_i$ where
$i=\{1,\ldots, N\}$. 
The Lagrangian contains kinetic terms,
dimension-less $\xi$ couplings to gravity, and a scalar potential $V$
\beq\label{eq:xi}
\Lag =\frac{(\partial_\mu \h_i)^2}{2} -V- \frac12 ( \bp^2+\xi_i  \h_i^2) R+\cdots\eeq
Here $\cdots$ denotes possible gauge vectors, fermions and  Yukawa interactions. 
The scalar potential is
\beq V = V_0 - \frac{M^2_{ij}}{2} \h_i \h_j - \frac{A_{ijk}}{3} \h_i \h_j \h_k + \frac{\lambda_{ijkl}}{4} \h_i \h_j \h_k \h_l 
=V_0+V_2+V_3+V_4
 .\eeq
An extra linear term is absent for those scalars $\h$ charged under gauge interactions,
and can be shifted away for the other neutral $\h$.
If some scalars are components of gauge multiplets, the gauge symmetries restrict their couplings.
We assume a mass scale $M$ somehow smaller than $\bp$.
As usual, we can rotate to a basis where $M^2$ is diagonal.\footnote{Furthermore, as long as we focus on counting minima,
without loss of generality we could rescale scalars 
to a basis where all $\h_i$ have the same $|M^2|$, at the  price of non-canonical kinetic terms.}

In gauge theories, self-cubics $A$ are allowed for scalars in
representations restricted by a trace-less condition:
for example the adjoint of SU groups, the symmetric of SO groups, the anti-symmetric of Sp groups.
Scalars in fundamentals have a unique self-quartic, and two inequivalent
self-quartics are
usually allowed for two-index representations (symmetric, anti-symmetric, adjoint).

\subsection{Independent scalars: the Lorentz$^N$ landscape}
Establishing if a generic potential has many local minima is a problem 
without general solutions.
One needs to proceed through direct enumeration:
find each extremum, expand around it and check
if all squared $N$ mass eigenvalues are positive.
Then the difficulty is that finding $10^{160}$ or more local minima is prohibitively slow.

In order to have a computable landscape, the authors of~\cite{NimaLand} 
considered a simple special potential
such that direct enumeration is not needed
to establish that it admits $M^N$ minima:
a ``rectangular''  potential given by the sum of
$N$ independent potentials $V_i$ with $M$ minima each,
\beq V(\h_1,\ldots ,\h_N)= \sum_{i=1}^N V_i(\h_i).\eeq
The action has an enhanced Lorentz$^N$ symmetry in the sub-Planckian limit.
In a renormalizable QFT one scalar has $M=2$ minima.
These minima give rise to a landscape of $M^N$ different values of the vacuum energies,
\beq \label{eq:Vrandom}
V= \min V +
\sum_{i=1}^N  \eta_i\, \Delta V_i  ,\qquad
\Delta V_i =V_i(\h_i^{\rm high}) - V_i(\h_i^{\rm low}) \ge 0
\eeq
where $\eta_i=0$ (1) selects the lower (higher) vacuum of $\h_i$.
The statistical distribution of $V$ tends to a Gaussian in the formal limit
$N\to\infty$ if a condition described by Lyapunov is satisfied.
Such condition roughly means that the $\Delta V_i$ must be comparable, in the sense that they do not
grow or decrease too much at large $N$.
A non-Gaussian landscape will be considered in section~\ref{nonGauss}.

\medskip

However, $N$ independent scalars do not give rise to a landscape of many different
 Higgs masses, unless the Higgs is added as a special extra
scalar that interacts with other scalars through mixed quartic or cubic couplings~\cite{NimaLand}.
Given that the Higgs mass now seems unnatural and can plausibly be
anthropically selected~\cite{1906.00986}, 
we would like to interpret its unnatural lightness by
finding a QFT landscape where the Higgs is a scalar with no ad-hoc special properties.
We then need to consider interacting scalars.

\subsection{Interacting scalars}
However, the presence of generic scalar interactions (cubic and/or quartic) 
can destroy the landscape of~\cite{NimaLand}, because
most stationary points become saddles rather than minima.
In the presence of large generic interactions one naively expects
that all $N$ mass squared eigenvalues are positive with $2^{-N}$ probability,
because  each eigenvalue can be either positive or negative at each minimum.
So one expects $(M/2)^N $ vacua, with
$M=2$ for a renormalizable potential.
Thereby a generic renormalizable potential has
$(M/2)^N \sim$ a few vacua, which means no landscape.
For example, just one  vacuum is present in the opposite ``elliptical'' limit
where the quartic has the form $(\sum_i c_i \h_i^2)^2$.

A more negative conclusion is reached applying
random matrix theory to the squared mass matrices of
the scalars at each extremum.
The probability that all eigenvalues are positive is smaller than $2^{-N}$ because
eigenvalues  repel.
A statistical argument implies that such probability falls off as $e^{-N^2/4}$ at large $N$
in the limit where all entries (diagonal and off-diagonal) are comparable~\cite{hep-th/0512050}.\footnote{In the regime where many minima become saddles with tachionic scalars,
the rare minima tend to have accidentally light scalars,
as large positive squared mass eigenvalues are even more statistically disfavoured
than positive squared mass.
Using random matrix techniques we estimate that  the lightest scalar is
accidentally lighter by an amount that scales as $N^{4/3}$.
Unless the number $N$ of scalars is huge, this factor is not significant for Higgs naturalness.}

This random matrix argument applies when off-diagonal cubics and/or quartics are large
enough as to make off-diagonal entries in the squared mass matrices $M^2_{ij}$ at the minima
so large that some eigenvalue becomes negative.
Scanning scalar masses (so that the Higgs can be accidentally light)
requires large enough
mixed interactions, possibly conflicting with having many minima.



\subsection{A bi-quadratic $\mathbb{Z}_2^N$ landscape}
In order to find a computable
landscape such that both the Higgs mass
and the vacuum energy density can be accidentally small,
we allow for possibly large mixed quartics that respect $\mathbb{Z}_2$ symmetries acting
independently on each scalar as $\h_i\to - \h_i$.
Each spontaneously broken $\mathbb{Z}_2$ gives rise to $2$ minima, 
such that $2^N$ minima arise when all $\mathbb{Z}_2$
are spontaneously broken.
The potential is written as
\beq V = \tilde V_0+V_2+V_4 + V_{\rm odd}\qquad\hbox{with}\qquad V_2=-\frac12 M_i^2 \h_i^2 \qquad\hbox{and}
\qquad V_4=\frac14 \lambda_{ij} \h_i^2 \h_j^2\eeq
where $V_{\rm odd}$ denotes extra smaller cubic or quartic terms that break the $\mathbb{Z}_2$ symmetries.\footnote{Various sources can generate a scalar potential with $\mathbb{Z}_2^N$ symmetry.
1) In the presence of gauge interactions with couplings $g$, one-loop RGE running towards low energy generates
positive quartics $\lambda_{ij} \circa{>} g^4/(4\pi)^2$ when gauge multiplets $i$ and $j$
are charged under a common gauge group.
2) In possible dimensionless theories of gravity, RGE gravitational
corrections at one-loop  generate mixed quartics
$\lambda_{ij} \circa{>} \ggr^4/(4\pi)^2$ 
among all multiplets, where $\ggr$ is a dimension-less gravitational coupling 
(see eq.~(23) of~\cite{1403.4226} for the precise RGE).
3) Special mixed quartics are obtained from the potential with independent scalars, 
taking into account that its Lorentz$^N$ symmetry is explicitly broken by the gravitational
$\xi$ couplings, assumed to be $\mathbb{Z}_2^N$ symmetric as in eq.\eq{xi}.
Rotating to the Einstein frame
$g_{\mu\nu}^{\rm Ein} = g_{\mu\nu}\times f$ with $f=1 + \xi_i \h_i^2/\bp^2$
the scalar kinetic term become non-canonical, and the Einstein-frame
potential becomes $V_{\rm Ein} = V/f^2$.
Expanding $V_{\rm Ein}$ in powers of $1/\bp$ gives mixed quartics
$6 \xi_i \xi_j \h_i^2 \h_j^2 V_0 /\bp^4$, thereby contributing as  a special
$\lambda_{ij}$ with rank 1.}
The potential is stable at large field values provided that all eigenvalues of $\lambda_{ij}$ are non-negative.
This relevant condition will play an extra role in the following.

In the $\mathbb{Z}_2$-symmetric limit the minimum conditions are
linear equations in $v_j^2=\langle \h_j\rangle^2$:
\beq M_i^2 =  \lambda_{ij} v_j^2 \label{eq:minVbiq}\eeq
which can be easily solved.
The potential can be re-written in terms of the $v_i^2$ as
\beq V= V_0+\frac14\lambda_{ij}(\h_i^2-v_i^2)(\h_j^2-v_j^2) + \cdots \qquad
V_0=\tilde V_0- \frac{M_i^2 (\lambda^{-1})_{ij} M_j^2}{4}
\label{eq:Vbiq}
\eeq
If some $v_i^2$ is negative the corresponding $\h_i$ has one minimum at 0
(for bounded-from-below potentials) and does not contribute to doubling the number of minima,
as the corresponding $\mathbb{Z}_2$ remains unbroken.

Let us here focus on scalars with $v_i^2>0$.
Their mass matrix is 
\beq V_{ij} = \frac{\partial^2 V}{\partial \h_i \partial \h_j}
=
(-M_i^2+ \lambda_{ik} \h_k^2) \delta_{ij} +2  \lambda_{ij}  \h_i \h_j=2  \lambda_{ij}  v_i v_j.\eeq
where the first term cancels at the minima. All minima have the same mass matrix of scalar fluctuations, up to
field redefinitions $\h_i\to - \h_i$.
Since $\det V = (\prod 2 v_i^2) \det\lambda$,
the squared masses eigenvalues of $V$ have the same sign as the
eigenvalues of $\lambda$.
This shows that all extrema are minima as long as the positivity condition 
$V_4\ge 0$ is satisfied.
The minimum condition (according to which all eigenvalues of $V_{ij}$ must be a positive at a minimum)
in general is restrictive and difficult to control.
Without special bi-quadratic potential it adds no extra condition.
Unlike in the more general situation considered in~\cite{hep-th/0512050} $V_{ij}$
is not a random matrix, and its positivity is guaranteed by positivity of the potential.

\medskip

Positivity of eigenvalues of $\lambda_{ij}$ does not imply that off-diagonal quartics must be small.
For example, for $N=2$ the positivity condition can be explicitly written as
\beq  \lambda_{11,22}\ge 0,\qquad \det\lambda=\lambda_{11}\lambda_{22}-\lambda_{12}^2 \ge0.\eeq
For generic $N$ we can consider, for example, the special structure
$\lambda_{ij}=\lambda  \, \epsilon_\lambda^{|i-j|}$ that
gives $\det\lambda = \lambda^N (1-\epsilon^2_\lambda)^{N-1}$,
so that positivity of $V$ 
is satisfied for $|\epsilon_\lambda|<1$. 
This $\lambda_{ij}$
leads to a Gaussian landscape of vacuum energies, and to a non-Gaussian landscape of Higgs masses of the type discussed in section~\ref{nonGauss}.

In general we can parameterize $\lambda_{ij}$ in terms of its eigenvalues $\lambda_i$ and
of its mixing matrix $R_{ij}$ as
\beq \label{eq:lambdaR}
 \lambda_{ij} = R^T\cdot  \diag (\lambda_i)\cdot R.\eeq
A Gaussian landscape for masses is obtained assuming comparable mixing angles.
The eigenvalues can be chosen independently of the
rotation matrix $R$, that can have large mixing angles.\footnote{
The rotation with largest mixing angles has $|R_{ij}|^2=1/N$
for $N$ odd 
(there are $N+1$ choices of signs such that $R^T \cdot R = 1$).}
Notice that $\lambda_{ij}$ can be rotated to a diagonal form 
by acting with  rotations $R$ on $\h_i^2$ (rather than on $\h_i$), 
an operation not allowed by the
symmetries of kinetic or mass terms.
This means that the bi-quadratic landscape is not equivalent to
the landscape of~\cite{NimaLand}.
Indeed, the $\mathbb{Z}_2^N$ symmetry of the bi-quadratic landscape
differs from the Lorentz$^N$ symmetry of the landscape of~\cite{NimaLand}.

\begin{figure}[t]
\begin{center}
\includegraphics[width=0.45\textwidth]{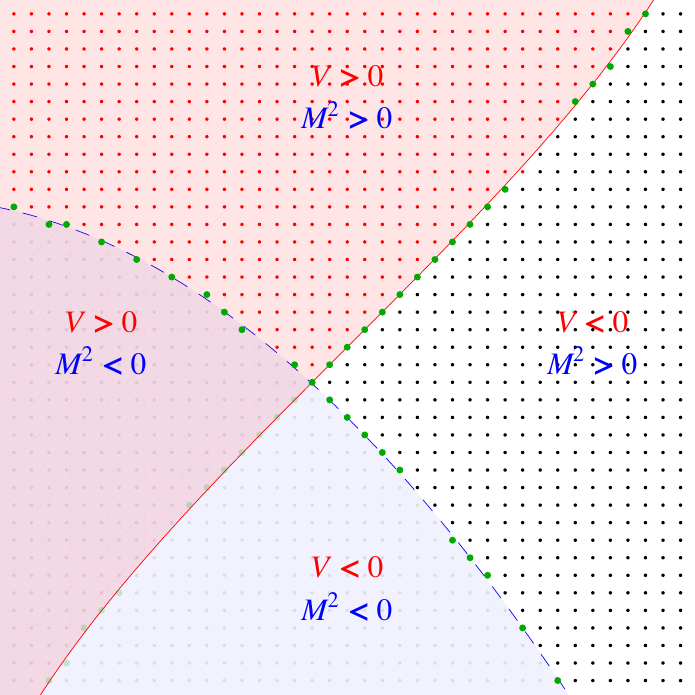}
\caption{\em Each dot is a minimum in the $\mathbb{Z}_2$-symmetric limit.
For illustrative purposes we replace
$2^N$ points in $N$ dimensions by points in 2 dimensions.
Adding $\mathbb{Z}_2$-breaking  terms some minima disappear or appear;
a light scalar is present around the boundary (blue curve);
the identity of the light  scalar (and thereby its Yukawa couplings) varies in the landscape.
Another boundary (in red) separates minima according  to the sign of their vacuum energy;
a small cosmological constant is present around the boundary.
\label{fig:LandscapePoints}}
\end{center}
\end{figure}

\subsection{An approximatively bi-quadratic landscape}
All vacua of the $\mathbb{Z}_2$ landscape have the same energy density and the same
scalar masses:
the $\mathbb{Z}_2$ symmetry cannot be exact
in order to have minima with accidentally small
vacuum energy and Higgs mass.
As long as the $\mathbb{Z}_2$-breaking effects $V_{\rm odd}$ are smaller than potential barriers
one still has a controllable landscape of many vacua.
Small linear terms and/or small cubic terms and/or
more generic quartics can contribute to $V_{\rm odd}$.

The $\mathbb{Z}_2$-breaking terms can be increased until
some scalar eigenvalues in some minima start becoming tachionic and
the corresponding $\mathbb{Z}_2$-symmetric minimum disappears.
If such boundary cuts the landscape in two comparable parts,
scalar masses are efficiently scanned so that
the observed light Higgs arises with its fine-tuning probability $\sim 10^{-34}$
and no extra suppression.

Among the surviving minima some have $V>0$, others $V<0$.
If the $V=0$ boundary cuts the landscape in two comparable parts,
the vacuum energy is efficiently scanned 
so that the observed vacuum energy arises with its fine-tuning probability $\sim 10^{-120}$
and no extra suppression.
These considerations are pictorially illustrated in fig.\fig{LandscapePoints}.

We considered the tree-level potential.
Quantum corrections at one and more loops  shift
the boundaries $V=0$ and $M^2=0$,
so that the  special vacua with small tree-level $V$ and $v$ acquire large $V$ and $v$.
What matters is that vacua with small $V$ and $v$ still exist,
as long as quantum corrections are small enough to preserve the
tree-level structure of the landscape.




%

\begin{figure}[t]
\begin{center}
\includegraphics[width=0.45\textwidth]{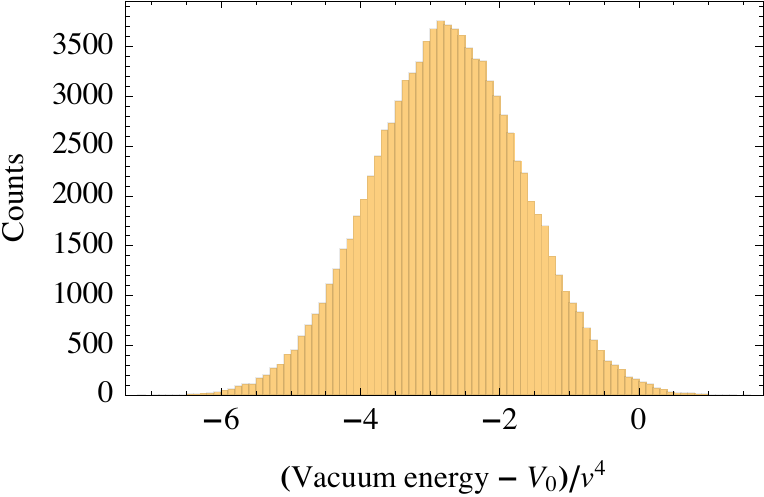}\qquad
\includegraphics[width=0.45\textwidth]{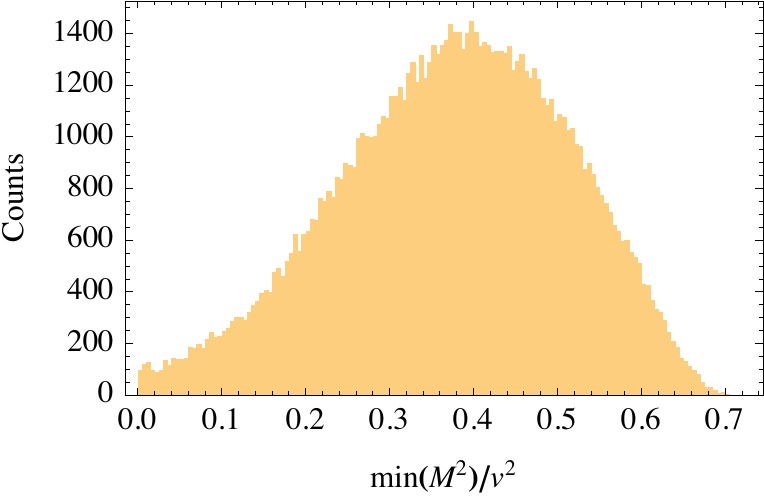}
\caption{\em We consider $N=100$ scalars with the
bi-quadratic potential of eq.\eq{Vbiq} for 
quartics $\lambda_{ij}$ as in eq.\eq{lambdaR}, with
comparable diagonal eigenvalues $\lambda_{i}\approx 1$ extracted from a Gaussian,
mixing angles $\theta\sim 0.1$, 
vacuum expectation values $v_i=v$.
We add cubic terms $V_{\rm odd}={\cal O}(0.1) v\h_i^3 $.
We compute $10^5$ random local minima and show their
distribution of the vacuum energy (left plot) and of the lightest scalar squared mass (right panel).
\label{fig:Landscape1}}
\end{center}
\end{figure}

\subsection{Scanning scalar masses and the vacuum energy}
We now estimate how the vacuum energy and the Higgs mass scan in the landscape.
We validate estimates by considering a numerical example
(with results  shown in fig.\fig{Landscape1}):
we consider a $\mathbb{Z}_2$ landscape with quartics as in eq.\eq{lambdaR},
assuming comparable quartic eigenvalues $\lambda_i \sim\lambda$, comparable masses $M_i\sim M$,
and comparable off-diagonal quartics that we 
can parameterize as  $\lambda_{\rm off}\sim \theta \lambda$
in terms of the angles $\theta\circa{<} 1$ of the rotation matrix $R$.
Under these assumptions we expect a Gaussian landscape.

The $\mathbb{Z}_2$ symmetries are broken by small cubics $-\frac13 A \h^3$ with
$A=2\epsilon M \sqrt{\lambda}$,
or by small linear terms $-\mu^3 \h$ with
$\mu^3=2\epsilon M^3 /\sqrt{\lambda}$, or by extra quartics.
For each scalar, its possible vacuum expectation values scan as
$\h \simeq  [\pm1 +\epsilon] M/\sqrt{\lambda} $.
Such vacuum expectation values contribute to vacuum energies as 
$ [-1 + {\cal O}(\epsilon) ]M^4/(4\lambda)$.
Summing the contribution of the $N$ scalars gives a distribution of values of $V$ 
and of squared scalar masses $m^2$ with standard deviation
\beq \label{eq:variations}
\delta V \approx \sqrt{N}\epsilon \frac{M^4}{\lambda},\qquad
\delta m^2 \approx \sqrt{N} \theta \epsilon M^2.\eeq 
A significant scanning of scalar masses, $\delta m^2 \sim M^2$
such that nearly massless scalars are not rare,
is obtained for \beq \sqrt{N}  \theta \epsilon\gtrsim 1.\eeq
This condition can be more easily satisfied in the $\mathbb{Z}_2^N$ landscape
for $\theta$ (i.e.\ off-diagonal quartics that do not split minima)
larger than $\epsilon$ (i.e.\ odd terms that split minima).

On the other hand, it is not guaranteed that the scanning of vacuum energies is wide enough to include
vacua with nearly vanishing vacuum energy.
This depends on the unknown
absolute value of the vacuum energy.\footnote{Its value can be guessed in different ways.
In string-like models with a heavy mass scale, the vacuum energy
might receive power-divergent quantum corrections.
In agravity-like models with no heavy mass scale,
the vacuum energy receives a precisely computable physical
 RGE correction of order $\ggr^4 \bp^4/(4\pi)^2$ (eq.~(54b) of~\cite{1705.03896}) where
$\ggr$ is a combination of the dimensionless gravitational coupling.
Furthermore, the vacuum energy receives a physical Quantum Field Theory
one loop RGE correction of order
${NM^4}/{(4\pi)^2}$.}
Many vacua with small $V$ are encountered if
the unknown overall additive constant in the potential is such that
$V=0$ lies around the common bottom of all minima in the $\mathbb{Z}_2$-symmetric limit,
as this is the region that gets most densely scanned in the presence of $\mathbb{Z}_2$-breaking terms.
No vacua with small $V$ are instead encountered if $V=0$ lies above the top of the potential at $\h_i \sim 0$ or much below the bottom of the minima.
Sizeable $\mathbb{Z}_2^N$-breaking terms, $\epsilon\circa{<}1$,
allow for a wider scan of the vacuum energy, 
up to a range of order $V_{\rm diff}=V_{\rm top} - V_{\rm bottom}\sim N M^4/\lambda$.
In view of this contribution, one can reasonably expect that vacua with $V=0$ are
$N_\sigma \sim \sqrt{N}$ standard deviations away from the central value of $V$ in the landscape~\cite{NimaLand}.
So the number of Minkowski vacua (with nearly vanishing vacuum energy) depends on $N$ as
$2^N e^{-N_\sigma^2/2} = e^{- (N2\ln 2 - N_\sigma^2)/2}$, which grows with $N$
only if $N_\sigma < \sqrt{N 2\ln 2} \approx 1.17 \sqrt{N}$.
This means that a generic landscape
can fail or succeed in generating many Minkowski vacua~\cite{NimaLand}.
Success is possible but not guaranteed.

\subsection{Scanning Yukawa couplings}
The authors of~\cite{NimaLand} assumed that the SM Higgs doublet $H$
is a special fixed scalar with Yukawa couplings to SM fermions $\psi$ described
 in a low-energy effective theory as $y(\phi/M) H\psi \psi$,
 where $y$ encodes non-renormalizable operators that depend on
landscape scalars $\phi$.
Then \cite{NimaLand} argued that the landscape can produce a distribution of Yukawa couplings $y$ peaked
at some value with small relative width $1/\sqrt{N}$, giving some predictivity.

\medskip

We consider a more general situation, where some 
linear combination $H = (h,0)$ of the many scalars $\h_i$
happens to be accidentally light in some vacua.
The composition of the light Higgs scalar can vary a lot in different vacua.
The full set of scalars can be decomposed as
\beq \h_i = \wp_i h +\hbox{(heavy scalars)}\eeq
with different $|\wp_i|<1$ in each vacuum.
Writing the Yukawa coupling of the scalar $\h_i$ as $y_i$,
the Yukawa coupling of the light higgs is given by
\beq \label{eq:Yuk}
y^{(n)} = \sum_i y_i \wp_i^{(n)}\eeq
where $n$ runs over the vacua where one Higgs doublet is accidentally light.
This structure is not captured by an effective field theory,
as the latter can only describe light degrees of freedom
omitting heavy ones, that instead play an important role.

Let us consider the sub-set of landscape scalars $\h_i$ with the same gauge quantum numbers as the Higgs doublet. 
The number $N_H$ of such scalars could be significantly smaller than $N$,
as most of the scalars $\h_i$ likely have different quantum numbers than the light Higgs $H$,
so that their $y_i$ vanish.


We experimentally know that
all such ultra-heavy Higgs doublets must have vanishing or small vacuum expectation values,
and thereby positive squared masses at the origin.
At the minima away from the origin, the mass matrix $M_H^2$
 of Higgs-like scalars 
gets affected by the vacuum expectation values of other landscape scalars.
The values of $M_H^2$ at each minimum
follow a statistical distribution that,
in many cases, tends to a multi-variate Gaussian centred on some $M_0^2$
with the variance of eq.\eq{variations}, so that
\beq M^2_H = M_0^2  \pm \sqrt{N} \theta \epsilon M^2.\eeq
An accidentally light Higgs-like scalar arises
along the boundary $\det  M^2_H=0$.
Whenever the variability is large enough that a light Higgs is not much rarer than its
fine-tuning factor $10^{-34}$, 
its composition encoded  in the $\wp_i^{(n)}$ coefficients vary a lot
(unless the fluctuations $\delta M^2$ are small enough, and the
central value $M_0^2$ is so skewed that a light Higgs is
dominantly found along a special direction that corresponds to an especially light eigenvector of $M_0^2$). 


The statistical distribution of $y$ can easily be non-Gaussian, for the following reason.
Given that Yukawas $y$ observed at low energy show large hierarchies, one can expect that the
same holds true for the values of the fundamental Yukawas $y_i$.
Then, even if $N_H$ were large, the sum over $N_H$ could not converge to a Gaussian.




\section{Vacuum decay in a landscape}\label{vac}
Vacuum decay in the landscape is not necessarily dominated by quantum tunnelling of a vacuum 
to the nearest vacuum.
Even considering one scalar, one can find sample potentials with $N>2$ minima
such that jumps over multiple minima either dominate the decay rate, or have zero rate:
the result depends on the relative height of minima and barriers
(see also~\cite{1108.0119}).

\smallskip

To start, we  consider the simple landscape of $N$ independent scalars.
We can compute the decay widths from a typical false vacuum, where
$N'\le  N$ of the scalars $\phi_i$ sit in their higher energy state.
Let us denote as $S_{i'}$ the bounce actions for the $N'$
transitions such that only one of such scalars
$\h_{i'}$  tunnels from its higher to its lower local minimum, with bounce $\h^{\rm b}_{i'}(r)$.
Naively,  these $N'$ bounces can be combined to give $2^{N'}$ bounces
where a generic sub-set of the $N'$ scalars tunnels.
By defining a tunnelling channel in terms of
$\eta_{i'}$ (equal to $1$ for those scalars that tunnel and to 0 for those scalars that do not tunnel)
one has an apparent multi-field bounce $\h^{\eta \rm b}_{i'}(r)= \eta_{i'} \h^{\rm b}_{i'}(r)$
with additive action $S_\eta = \sum_{i'=1}^{N'} \eta_{i'} S_{i'}$.
This would mean that multi-field bounces negligibly contribute to the vacuum
decay rate, despite their large number $2^{N'}$.
A similar sharper conclusion is reached observing
that multi-field configurations are not true bounces,
because true bounces  have only one negative mode~\cite{Coleman:1987rm},
while multi-field configurations have $\sum \eta_{i'} > 1$
negative modes.
So the number of bounces is $N' \ll 2^{N'}$.\footnote{We thank J.R.\ Epinosa for discussions about vacuum decay.}

\medskip

\begin{figure}[t]
\begin{center}
$$\includegraphics[width=0.44\textwidth]{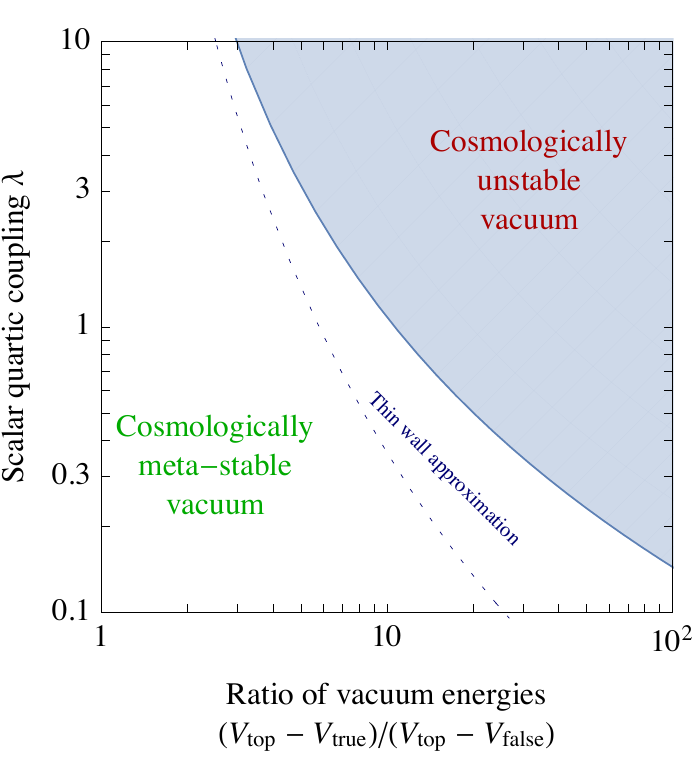}
\qquad\raisebox{5mm}{\includegraphics[width=0.44\textwidth]{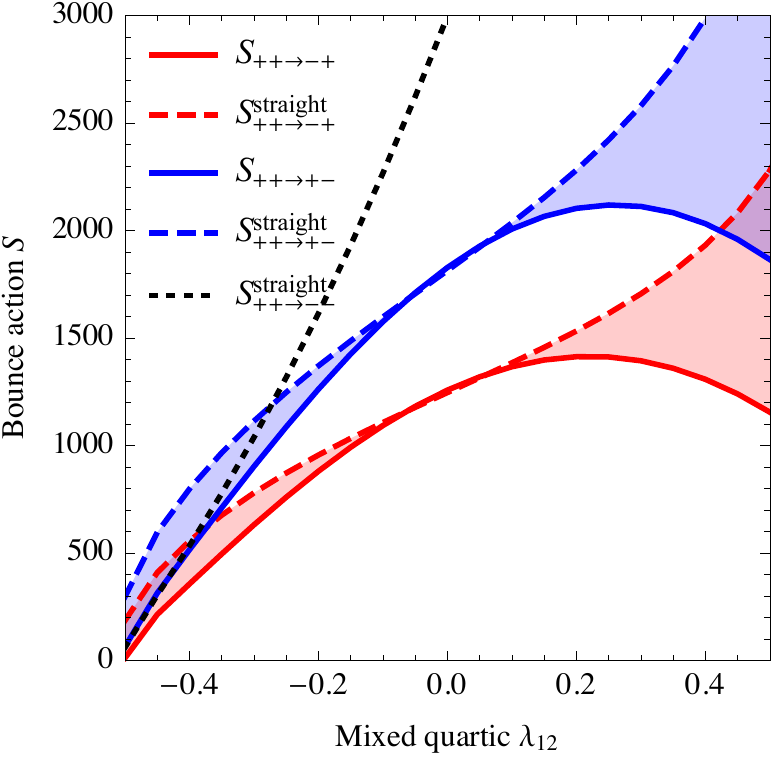}}
$$
\caption{\em {\bf Left}: we consider a potential $V = -  M^2 \h^2/2 - A \h^3/3 + \lambda \h^4/4$
and show the largest $\lambda$ allowed by vacuum stability as function of
the ratio $R=(V_{\rm top}-V_{\rm true})/(V_{\rm top}-V_{\rm false})$
 of the vacuum energies at the false and true vacua.
We see that even large non-perturbative $\lambda$ are allowed for order one ratios
$R\sim 2$.
{\bf Right}: bounce actions for a 2-field potential
with $\lambda_{11}=\lambda_{22}=1$, $v=v_1=v_2$,
$V_{\rm odd} = 0.2\h_1 v_1^3+ 0.18 \h_2v_2^3$.
\label{fig:VacuumDecayBound}}
\end{center}
\end{figure}

Let us compute their actions $S\equiv S_{i'}$ corresponding to a tunnelling of $\phi \equiv \phi_{i'}$ only.
The potential along $\h$ has the form 
$V(\h) = V_0 - \sfrac{M^2\h^2 }{2}  - \sfrac{A \h^3}{3}  + \sfrac{\lambda \h^4}{4} $.
The bounce action $S$ 
can be computed
by rescaling to dimensionless variables $x=\tilde{x} \times {6\sqrt{\lambda}} /[{A}(1-3u)]$
and  $\h=A(1-u)/2\lambda + \tilde \h \times A (1-3u)/6\lambda$,
shifting the field such that $\h=0$ becomes the false minimum\footnote{A constant term has been omitted from the potential, as it is irrelevant at $M \ll \bp$.
Gravitational corrections depend on the absolute heights of the potentials 
making vacuum decay from Minkowski slower at sub-Planckian energies~\cite{1706.00792,1808.00420}.}
and rotating to Euclidean time, obtaining~\cite{Adams:1993zs}
\beq \label{eq:S4}
S= \frac{1}{\lambda} 
\int d^4\tilde{x}_E\left[\frac{(\partial \tilde{\h}_{\rm b})^2}{2}+
\delta \frac{\tilde{\h}_{\rm b}^2}{2} - \tilde{\h}_{\rm b}^3 +
\frac{\tilde{\h}_{\rm b}^4}{4}+\tilde{V}_{\rm false}\right]
= \frac{4\pi^2}{3\lambda(2-\delta)^3}\times (13.8\delta-10.8 \delta^2+2.1 \delta^3)
\eeq
where the first term is the  thin-wall limit, and
the second term (equal to $1$ for $\delta=2$)
approximates the full result, as found by numerically computing the 
bounce solution $\tilde{\h}_{\rm b}$.
The parameter $\delta$ equals
$ \delta = {18u(u-1)}/{(1-3{u})^2}$ where $ u = \sqrt{1 + 4\lambda{M^2}/{A^2}}$
and ranges between $\delta=0$ (two vacua with very different energies) and $\delta=2$ (two degenerate vacua).
Cosmological stability demands $S \circa{>} 530$. 
The resulting upper bound on $\lambda$ is plotted 
in fig.\fig{VacuumDecayBound} as function of 
\beq R \equiv \frac{V_{\rm top}-V_{\rm true}}{V_{\rm top}-V_{\rm false}}=
\frac{8 (9-4 \delta )^{3/2}}{\left(\sqrt{9-4 \delta }-3\right)^2 \left(2 \delta +\sqrt{9-4 \delta }-3\right)} \ge 1
\eeq
 where
$V_{\rm top}$ is the potential at the top of its barrier.
A significant scanning of the vacuum energy requires somehow splitted minima,
$R\circa{>}2$ i.e.\ $\delta\circa{>}1.9$,
and thereby $\lambda\circa{<}10$ to have cosmological stability.
A quartic $\lambda=1$ allows cosmological stability of 
two vacua up to $R \circa{<} 10$.

The total decay  rate is found summing over all decay channels,
 $e^{-S_{\rm eff}}\equiv \sum_{i'} e^{-S_{i'}}$,
 and tends to be dominated by the  single tunnelling with lower bounce action,
\beq S_{\rm eff} \approx \min_{i'} S_{i'}.\eeq 
The authors of~\cite{NimaLand} argued that cosmologically long-lived vacua
demand relatively small quartic couplings $\lambda\circa{<}0.5$
because the Euclidean bounce action that controls vacuum decay rates
$d^4\wp/d^4x \sim M^4 e^{-S}$ was estimated as $S \sim 27\pi^2/\lambda$.
Cosmological stability demands $S \circa{>}4 \ln M/H_0 \sim 530$ 
where $H_0$ is the present Hubble rate and $M\sim \bp/10$ the typical scalar mass.
The authors of~\cite{NimaLand} argued that such a small quartic might
conflict with the requirement of generating a wide enough scan
of vacuum energies, within some assumptions about the potential parameters.
Our  estimate of the generic vacuum decay bound is
weaker than in~\cite{NimaLand} mostly because we included the $(2-\delta)^{-3}$ factor:
as a result we find a weaker bound $\lambda{\circa{<}}10$.

\medskip




Let us next consider a landscape of interacting scalars.
As long as mixed interactions are small enough
the situation remains  as in the landscape of independent scalars:
only $N'$ vacuum decay channels are open out of $2^{N'}$.
Their bounce actions $S_{i'}$ however must now be computed taking all scalars into account,
as tunnelling no longer proceeds along straight trajectories
from one minimum to a nearby minimum.
Computing the action along a straight trajectory only gives an 
upper bound $S^{\rm straight}$ on the tunneling action.\footnote{$S^{\rm straight}$ can be computed as follows.
The straight trajectory from one generic minimum with $\langle \h_i\rangle=w_i$ 
to another generic minimum with $\langle \h_i\rangle=w'_i$ can be 
 parameterized by a canonical scalar combination $\h$ given by 
\beq \h_i = w_i+\h n_i,\qquad n_i \equiv \frac{w'_i-w_i}{v},\qquad
v\equiv \sqrt{(w-w')\cdot(w-w')}\eeq
where $n_i$ is a unit versor.
The action $S^{\rm straight}$ is then given by eq.\eq{S4} with $\lambda = \sum \lambda_{ij} n_i^2 n_j^2$.}
Numerical codes suggest that, for small and 
negative off-diagonal quartics, tunnelling trajectories prefer to circle `inside' (closer to the origin) giving lower bounce actions $S_{i'}$ with  respect to what obtained setting the cross-quartics to zero.
If off-diagonal quartics are small and positive, tunnelling trajectories prefer to circle `outside',
giving higher bounce actions $S_{i'}$.
The right panel of fig.\fig{VacuumDecayBound} shows a numerical example.
Tunnelling to non-nearby minima can appear when interactions are so  large that they
risk to destroy the landscape.

\subsubsection*{Vacuum decay from SM-like minima with a light scalar}
We must next  consider the possibility that
vacuum decay could be catastrophically faster for those 
special minima of physical interest
where one scalar $h$ is accidentally light,
so that its mass term negligibly contributes to the potential barrier.
A scalar can become very light in two different ways:
\begin{itemize}
\item {\bf Light scalars with a large vacuum expectation value} $v\sim M$,
that thereby cannot be identified with the Higgs.
For example, we consider the $\mathbb{Z}_2$ landscape.
Adding $\mathbb{Z}_2$-breaking terms $V_{\rm odd}$
some of the vacua present in the $\mathbb{Z}_2$-symmetric limit get destabilised (reducing the landscape), others persist:
vacua around the boundary between these possibilities feature a light scalar.
However, such vacua have very small potential barriers
and thereby their vacuum decay is catastrophically fast.\footnote{This phenomenon can be illustrated considering
one scalar $\h$
with renormalizable potential $V=M^2 \h^2/2+A\h^3/3+\lambda \h^4/4$:
the same tuning of the parameters that makes $V''$ especially small at one minimum 
(relatively to the other minimum) also gives a stronger suppression of the potential barrier.}


\item {\bf Light scalars with small or vanishing vacuum expectation value},
that can be identified with the Higgs.
Scalars $\h$ with a large positive squared mass at the origin $\phi_i=0$
do not contribute to the landscape in the $\mathbb{Z}_2$-symmetric limit. 
Adding $\mathbb{Z}_2$-breaking terms, 
vacuum expectation values of other scalars induce a landscape of scalar masses 
for scalars with no vacuum expectation values in the $\mathbb{Z}_2$-symmetric limit.
As a result, the origin ceases to be a minimum for some scalars that get destabilised
(enlarging the landscape), others persist around the origin:
vacua around the boundary between these possibilities feature a light scalar.
Vacuum decay is again catastrophically fast if such light scalar has a very large cubic-self coupling $A$.\footnote{Ignoring non-perturbative effects
(that might limit the possibility of tuning $m\ll A$)
such scalars acquire at tree level vacuum expectation values $v \simeq - m^2/3A$ if $m^2<0$.}
\end{itemize}
Cubics are expected to be of order of the heavy scale $M$, 
unless suppressed by accidental tunings (we ignore this possibility) or theoretical reasons.
For scalars at the origin, one possible theoretical reason is
group theory of some unbroken gauge invariance.
Scalars with self-cubics forbidden by gauge invariance
can be accidentally light without being
accompanied by catastrophically fast vacuum decay,
and can acquire a small vacuum expectation value of order $m/\sqrt{\lambda}$.

This is what happens for the Higgs doublet $H = (h,0)/\sqrt{2}$: a self-cubic $H^3$ or $H^2 H^*$ is forbidden
by $\SU(2)_L$ and/or ${\rm U}(1)_Y$ invariance.
The SM minimum with nearly vanishing $h=v\ll M$
lies near to the special gauge-invariant point in field space $h=0$,
so that the SM minimum is surrounded in all directions
by a potential barrier $\lambda_H |H|^4$
(as well as by potential barriers of other scalars involved in the tunnelling).
The self-quartic of the SM Higgs equals $\lambda_H \approx 0.126$
at the weak-scale.\footnote{And undergoes RGE running possibly becoming small and negative around the Planck scale or some orders of magnitude below.
This instability of the SM potential does not induce
catastrophically fast vacuum decay~\cite{hep-ph/0104016,1307.3536}.}

\begin{figure}[t]
\begin{center}
$$\includegraphics[width=0.45\textwidth]{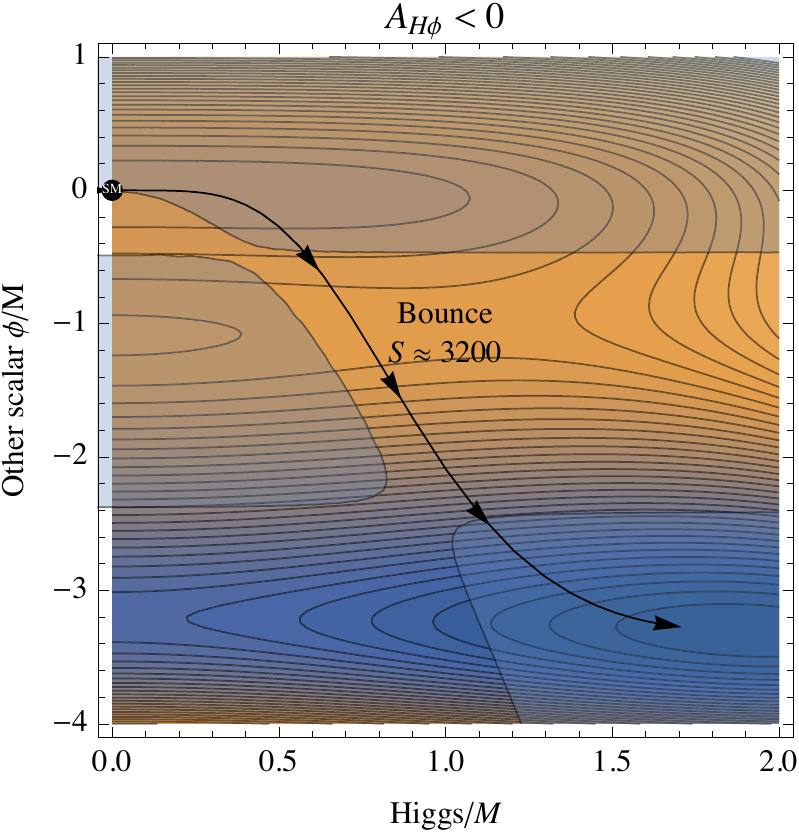}\qquad
\includegraphics[width=0.45\textwidth]{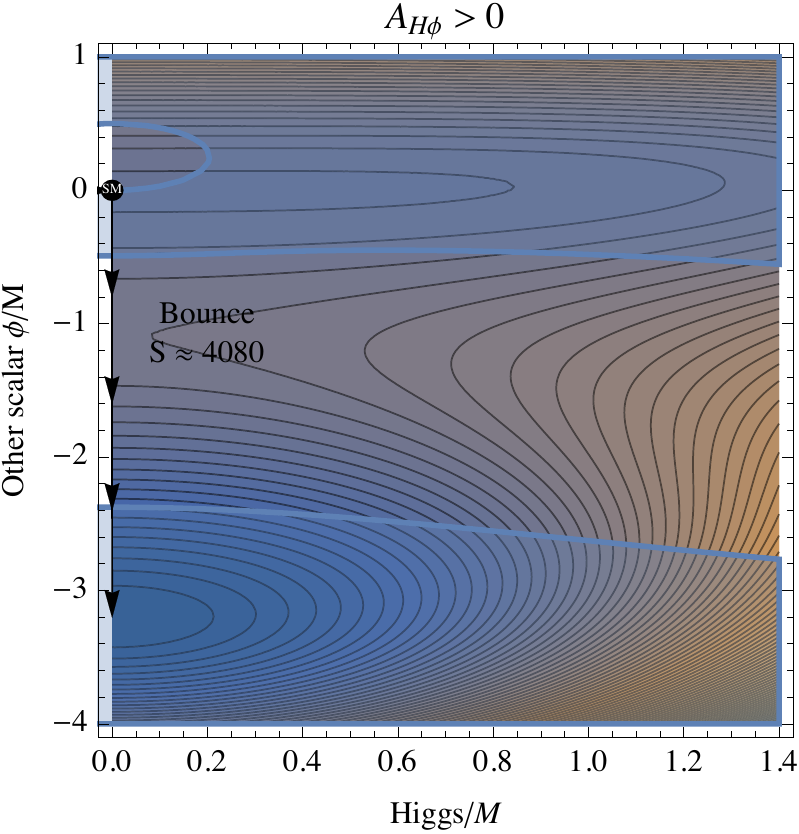}$$
\caption{\em Numerical examples for the two
characteristic classes of landscape potentials of eq.\eq{VhS}
that lead to a light Higgs-like scalar and another heavy scalar $S$.
The SM false vacuum lies at the origin.
Shaded regions have $\det V_{ij}>0$.
In the left (right) panel the true vacuum has $h\sim M$ ($h\ll M$).
The curve with arrow indicates the bounce trajectory;
the numerical values of the bounce action show that 
Higgs-like scalars can be light without leading to catastrophically fast
vacuum decay. 
\label{fig:VacuumDecaySM}}
\end{center}
\end{figure}

While this effective field theory argument captures the key point,
vacuum decay cannot be approximated by integrating out heavier fields
and obtaining an effective field theory in terms of $h$ only, 
because the range of validity of such Taylor-like expansion
is not enough to cover the ${\cal O}(M)$ field values relevant for vacuum decay.
We thereby study what happens in the presence of one extra landscape scalar $\h$ with  a far-away minimum.
If $\h$ is a singlet under  SM gauge interactions we can
shift $\h$ such that the SM-like minimum corresponds to $\h=0$ and the most generic
 potential can be written as
\beq \label{eq:VhS}
V(h,\h) = \frac{\lambda_H}{4} (h^2 - v^2  )^2 +\frac{M^2}{2} \h^2 
-\frac{A_{H\h}}{2} \h h^2 
- \frac{A_\h}{3}\h^3+
 \frac{\lambda_{H\h}}{4}h^2 \h^2 + \frac{\lambda_\h}{4}\h^4.\eeq
Fig.\fig{VacuumDecaySM} shows numerical examples of $V$
in two characteristic situations.

In the left panel we assume $A_{H\h}<0$ so that the weak symmetry 
is more strongly broken in the true vacuum than  in the physical false vacuum.
The bounce involves vacuum expectation values of order $M$
for the Higgs field, 
so that the Higgs quartic potential provides a significant potential barrier,
together with the $\h$-dependent part of the potential.
As a consequence
vacuum decay can be extremely slow for typical values of the parameters.
This is confirmed by the numerical example in the figure.\footnote{The bounce actions and trajectories are easily computed implementing the method of~\cite{1805.03680,Espinosa:2018szu}.
Our results agree
with results obtained running the public code of~\cite{1908.10868}, based on a different method.}
The bounce action negligibly depends on the weak scale $v$ in the limit $v \ll M$.

In the right panel we assume $A_{H\h}>0$ so that the weak symmetry is
preserved in the true vacuum.
In the limit $v\ll M$ vacuum decay proceeds purely through $\h$, with $h$ sitting at zero,
so that its quartic potential does not contribute to the barrier,
and the small Higgs mass does not facilitate vacuum decay:
its vacuum decay rate is not influenced by the lightness of $h$.
Vacuum decay can be fast in an unfavourable landscape, 
but the accidental
lightness of the Higgs boson is not a risk factor that implies catastrophically fast vacuum decay.

Qualitatively new potentials arise in the presence of heavy scalars charged under the SM gauge  group:
one can have mass mixings $\mu^2 \,H H^{\prime *}$ with extra heavy Higgs doublets $H'$,
and cubics $A\,H H^{\prime *} \h $, $A H H' \h$ with extra heavy scalars $H'$ and $\h$ 
(for example they could be an extra doublet $H'$ and an extra singlet $\h$;
SU(5) unification models predict
Higgs color triplet $H'$ and a bi-fundamentas $\h$ under $\SU(2)_L\otimes\SU(3)_c$).
Mass mixings can be ignored, as they vanish in the basis where
$H$ is the accidentally light scalar.
For the same reason one can assume that $\med{\h}=0$ around the SM minimum,
so that the new cubics are harmless. 


We thereby conclude that a vacuum with a light scalar is not more unstable, 
provided that the light scalar (like the SM Higgs)
is charged under gauge interactions that forbid self-cubics.

\section{A landscape in dimension-less theories?}\label{CW}
The Higgs mass naturalness issue is avoided if
the Higgs is not significantly coupled to new physics much
heavier than the weak scale, given that the squared Higgs mass
would not receive unnaturally large physical quantum corrections.
This possibility is consistent with experiments.

Furthermore, in the context  of dimension-less theories,
the small weak scale can be dynamically 
generated \`a la Coleman-Weinberg. 
Many models of this type have  been recently considered,
using either small or non-perturbative interactions.
In particular, 
a possible way of including gravity and the Planck scale was discussed in~\cite{1403.4226}.
However, the smallness of the vacuum energy $V$ is unnatural in such context because we know that gravity is coupled to the  Standard Model,
so that $V$ receives unnaturally large  corrections of order $v^4$.
Furthermore, dynamical generation of mass scales in dimension-less theories
typically leads to a large negative $V$.

We here show that  
dimension-less theories can give rise to a landscape of vacua,
including vacua with a vanishingly small cosmological constant.
Indeed, when some first scalar $\h$ dynamically
acquires a vacuum expectation value or a condensate, 
the dimension-less potential of all $\h_i$
becomes a potential with massive parameters,
that can give rise to a landscape of vacua  in the  same way discussed earlier.
 
Concerning the sign of the vacuum energy,
a Coleman-Weinberg potential
$V(\h) \approx \lambda(\h) \h^4/4$
with logarithmic running of the quartic, $\lambda(\h) \approx \beta_\lambda \ln \h/\h_*$,
has a minimum at $\phi\approx \phi_* e^{-1/4}$ with $V< 0$.
More in general, since $V(0)=0$, any global minimum at $\phi\neq 0$
necessarily has negative
vacuum energy.
However, Coleman-Weinberg potentials can also have local minima 
with $V>0$ provided that the quartic $\lambda$ runs in an appropriate tuned way,
discussed for example in~\cite{1403.4226}.
In a landscape with enough scalars  and vacua, the tuning needed for
the appropriate RGE running of $\lambda$ is the same tuning needed for a vacuum with anthropically
selected small vacuum energy.

In conclusion, dimension-less theories are not incompatible
with a landscape.

\section{Non-Gaussian landscapes and predictivity}\label{nonGauss}
A theory able of predicting the values of the `fundamental' constants in its
low-energy QFT limit can remain untestable in sub-Planckian experiments
if the number of its vacua is too large.  
The smallness of the  cosmological constant and of  the weak scale naively suggest
a landscape of at least $10^{123+35}\sim 10^{160}$ vacua.
On the other hand,
SM parameters have been measured so far up to about 72 digits of precision, 
in the sense that all experimental information can be condensed in
about 72 
dits (``bits'' in base 10) of information:
\begin{itemize}
\item Measurements of the  gauge couplings $g_{1,2,3}$
amount to about 14 digits of precision;
\item The Higgs quartic $\lambda_H$ is known from $M_h$ and $v$ with about 2 digits of precision.

\item Yukawa couplings of leptons are known to about 18 digits.

\item Yukawa couplings of quarks are known to about 8 digits.

\item The CKM mixing angles among quarks  add about $6-7$ extra digits.
QCD uncertainties limit the accuracy that can be reached measuring light quarks.

\item Neutrino masses and their mixings add about 8 extra digits.

\item Cosmology provides about 6 extra digits of precision
(and calls for a Dark Matter  extension of the SM).

\item Very small ratios among scales ($V/M_{\rm Pl}^4$ and $v^2/M_{\rm Pl}^2$)
and bounds (in particular on the QCD angle) give about 10 extra digits,
depending on how zeroes are counted.
Furthermore, one can add the number of chiral fermion generations  and
the dimension of gauge groups to the list of `coordinates'  that define the SM.

\end{itemize}
Thereby a landscape of more than $10^{72}$ vacua risks being untestable.
In the worst case where landscape predictions are distributed
with a feature-less Gaussian-like distribution,
vacua compatible  with all what  we observed can exist a huge number of times just  by pure change.
Narrow Gaussians offer increased predictivity~\cite{NimaLand}, 
but this hope conflicts with the observation that
$m_e$, $m_u$, $m_d$ are small and seem anthropically selected.

\begin{figure}[t]
\begin{center}
$$\includegraphics[width=0.44\textwidth]{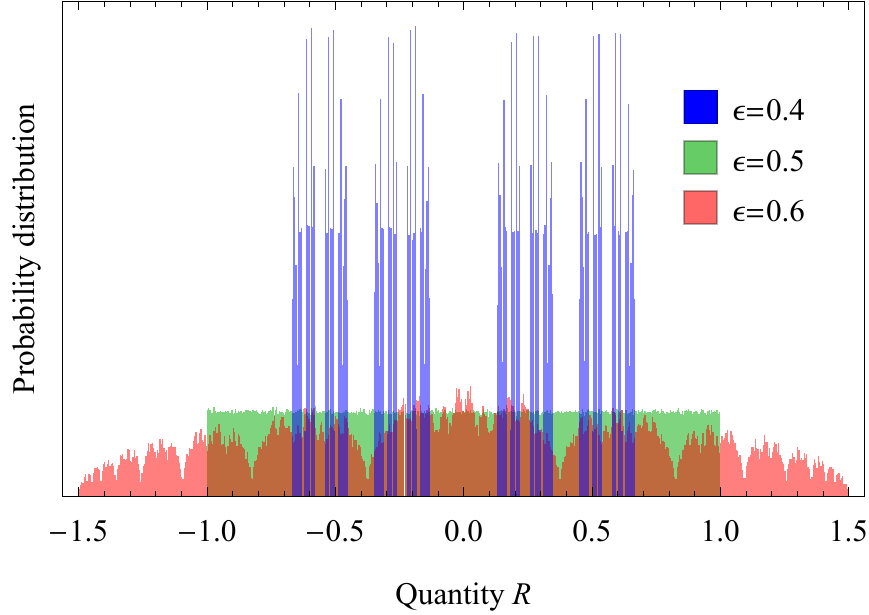}\qquad
\includegraphics[width=0.48\textwidth]{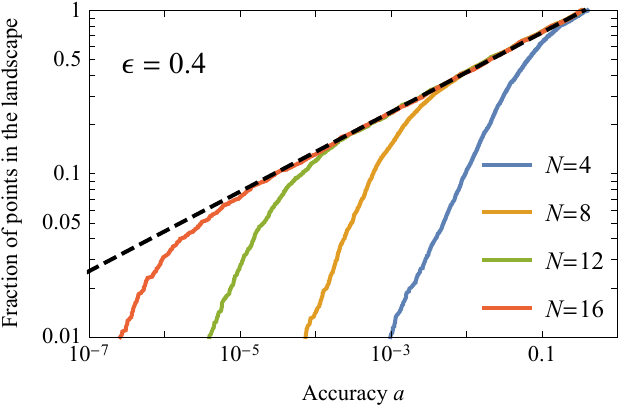}$$
\caption{\em {\bf Left:} distributions of a generic quantity $R$
given by the landscape of eq.\eq{R} at large  $N$ and for $\epsilon<1/2$ (blue fractal),
$\epsilon=1/2$ (green flat), $\epsilon>1/2$ (red  non smooth). 
A wide Gaussian would be obtained  for $\epsilon=1$.
{\bf Right}: fraction of the parameter space covered 
in fractal landscapes for different $N$.
\label{fig:fractal}}
\end{center}
\end{figure}


We then discuss a possible  non-Gaussian landscape.
Let us consider a quantity $R$ that (analogously to the vacuum energy in eq.\eq{Vrandom})
takes possible values
\beq  \label{eq:R}R=\sum_{n=1}^N r_n  f_n \qquad\hbox{with} \qquad
r_n = \pm 1 .\eeq
Viewing $R$ as a sum of random variables with values $\pm f_n$ 
and variance $\sigma_n^2 = 2 f_n^2$,
the Lyapunov central-limit theorem tells that the statistical
distribution of $R$ converges for large $N$ to a Gaussian 
if all $f_n$ are comparable.
We here explore what happens when this condition is violated, 
considering for example the case 
$f_n=\epsilon^n$ where $\epsilon$ is a possibly small constant.
Similar results would be obtained in the more general case $f_n \approx \epsilon^n$.

This example might apply to cases of physical interest: $R$ could be a Yukawa coupling, or 
a scalar squared mass in a landscape with mixed scalar quartics $\lambda_{ij} \propto \epsilon^{|i-j|}$
(or, more simply, a landscape where scalars do not have a common mass scale $M$).
Taking into account that $\epsilon$ and $1/\epsilon$ give the same distribution up to a rescaling,
we can focus on $\epsilon\le 1$, finding:
\begin{itemize}
\item For $\epsilon <1/2$ 
the distribution of $R$ shows a fractal structure of many separated peaks,
as illustrated  in fig.\fig{fractal}a.
A random number, defined up to an accuracy $a=\epsilon^N$ 
equal to the smallest contribution to $R$,
is present in the fractal landscape with probability $\wp=(2\epsilon)^N$
(black dashed line in fig.\fig{fractal}b; the other curves show
the result for different accuracies $a$).
For $N\to\infty$ one gets a fractal with dimension  $d=\ln2/\ln(1/\epsilon)$.
As large $N$ is needed, the probability $\wp$ becomes quickly small for $\epsilon < 1/2$.
For example,
the Higgs squared mass should be vanishing up to an accuracy of about
$a \approx M_h^2/M_{\rm Pl}^2\sim 10^{-36}$ in a Planck-scale landscape,
such that at least $N = \ln a/\ln\epsilon \approx 83/|\ln\epsilon|$ scalars are needed.
The probability that a light Higgs is present in the fractal landscape 
is already smaller than $1\%$ for $\epsilon =0.48$.

\item
For $\epsilon=1/2$ 
one gets a flat landscape distribution that fills the range $-1\le R\le 1$
because $R$ is like writing a random number in base 2.
This feature-less landscape is successful, 
provided that the scanned range contains the desired value
(flat space, light Higgs, etc).

\item For $1/2<\epsilon<1$ one gets successful landscape distributions that 
fill the range $|R|\le \epsilon/(1-\epsilon)$,
while still exhibiting significant sub-structures,
as illustrated  in fig.\fig{fractal}.
When computing multiple quantities $R$, one can hope for a
peaked enough multi-dimensional probability distribution such that 
some predictions are possible.

\item 
For $\epsilon=1$ the sub-structures disappear leaving a successful but feature-less
Gaussian landscape, 
where finding flat  vacua needs brute-force computations~\cite{hep-th/0602072}.
\end{itemize}



\section{Conclusions}\label{Conclusions}
Theorists developed beautiful theories of new physics such that
the weak scale would  naturally  be
much smaller than the Planck scale.
Such theories 
are now in trouble with bounds on new physics:
Nature rejected these theories up to tunings of a part in 100-1000.
Such theories remain less tuned than the SM up to the Planck scale,
but the absolute  level of  tuning might mean that nature
is not a contest where the  relatively less tuned theory wins:
our ideas of naturalness might be missing some much bigger ingredient.

The two failures of naturalness with the  
weak scale $v$ and with the vacuum energy $V$  
prompts us to consider alternative ideas.
Anthropic selection in a landscape of many vacua 
(also indicated by the special values of light fermion masses)
seems the most plausible  interpretation of the small values of $v$ and $V$.

We presented a renormalizable Quantum Field Theory  potential of $N\sim\hbox{few hundred}$ scalars that leads to an efficient scanning of the vacuum energy as well
as of the weak scale.  We avoid introducing any ad hoc structure for the Higgs.
In general, scalars with accidentally  small masses $m \ll M_{\rm Pl}$
can arise in a landscape where scalar masses are widely scanned.
But this risks removing almost all vacua from the landscape, 
as $m^2=0$ is the critical point where minima (all $m^2>0$) are lost becoming
saddle points (some $m^2<0$).

This potential problem cannot be controlled if this needs counting $\gg 10^{160}$ minima.  
We bypassed this practical difficulty by assuming an appropriate 
form for the scalar potential, such that vacua break approximate $\mathbb{Z}_2^N$ symmetries.
Furthermore, in section~\ref{CW} we found that a successful landscape for $v$ and $V$
remains possible also within dimension-less theories 
where masses are dynamically generated \`a la Coleman-Weinberg.

\smallskip

In section~\ref{vac} we  considered vacuum decay, finding that enough stability is possible,
even for relatively large quartics $\lambda\sim 1$ (see fig.\fig{VacuumDecayBound}).
Furthermore, we found that vacua with an accidentally light scalar
avoid catastrophically fast vacuum decay provided that the light scalar has
no self-cubic.  The SM Higgs doublet satisfies this property thanks to electroweak gauge invariance.

The accidentally light Higgs doublet present in rare vacua
is a different combination 
of the various weak doublets present in the  full theory: thereby 
it has different Yukawa couplings in different vacua, allowing 
for anthropic selection of light fermion masses, as suggested by data.

Finally, in section~\ref{nonGauss} we went beyond previous studies that
assumed values of the landscape parameters that result in Gaussian probability distributions
for the various quantities of interest.  
We found that in simple cases one can instead obtain fractal distributions
as well as less  dramatic
continuous distributions with peaked sub-structures.

While possibly being the correct physics,
a landscape of many vacua risks being vacuous physics,
as long as it does not provide testable implications.
In section~\ref{nonGauss} we tried to quantify the amount
of information measured so far at low energy comparing it to
the  `entropy' of the landscape.
Landscape distributions with narrow peaks would
have less `entropy' than a Gaussian landscape, and would thereby
ruin relatively less  the predictions  of a high-energy theory,
that might remain testable in low-energy experiments.

\subsubsection*{Acknowledgements}
This work was supported by the ERC grant NEO-NAT.

%
%


\end{document}